\documentstyle[12pt]{article}
\newcommand{\be}{\begin{equation}}
\newcommand{\ee}{\end{equation}}
\newcommand{\bea}{\begin{eqnarray}}
\newcommand{\eea}{\end{eqnarray}}

\begin{document}
 \begin{titlepage}

\begin{flushright}
CERN-TH.6916/93
\end{flushright}
\vspace{20 mm}

\begin{center}
{\huge A Matrix-Model Black Hole}
\end{center}

\vspace{10 mm}

\begin{center}
Ulf H. Danielsson\\
Theory Division, CERN, CH-1211 Geneva 23, Switzerland.

\end{center}

\vspace{2cm}

\begin{center}
{\large Abstract}
\end{center}
In this paper the $c=1$ matrix model deformed by a $1/x^{2}$ piece
is discussed. Tachyon correlation functions are calculated up to
genus two using methods similar to those for
the undeformed case.
The possible connection with the two dimensional black hole
is also considered. In particular, restrictions on naked singularities
imposed by the matrix model are found.

\vspace{3cm}
\begin{flushleft}
CERN-TH.6916/93 \\
June 1993
\end{flushleft}
 \end{titlepage}
\newpage

\section{Introduction}

It has turned out
to be remarkably difficult to find a matrix model
describing the two dimensional black hole. The reason is our
poor understanding of the target space physics in the standard
$c=1$ matrix model, thought to describe the linear dilaton vacuum.
If we had such a model, we could address the many questions raised
by the study of two dimensional black holes using field theory
\cite{CGSH}. There, severe ambiguities were discovered which have
made progress difficult, see e.g. \cite{harstro,stro} for reviews.
String theory in the form of matrix models
might possibly suggest some answers.

There is already a string theoretic black hole, the one
of \cite{wit}. It has been solved to all orders in $\alpha '$ on the
sphere. Higher genus, however, is very complicated to handle in
this field theory approach. A matrix model method, if it existed,
would be much preferable.
The genus expansion is also an expansion in $\hbar$
and therefore
essential for understanding the quantum black hole.

A lot of the confusion in the field has been due to an incorrect
identification of the Liouville mode in the matrix model. It
was tempting to identify the time-of-flight variable in the
inverted potential as the Liouville mode. This is natural from
the point of view of collective field theory, but not true,
even approximately. In fact, as was shown in \cite{moore,mose}, the true
identification is through the loop operator by an integral
transform. This gives an {\it explicit} construction of the
Wheeler-de-Witt equation, which coincides with the mini super
space Wheeler-de-Witt equation obtained from field theory.

There has been several
attempts to derive matrix model black holes
by working with spatial coordinates simply related to the
time of flight \cite{yang,russo}. In this connection
the $1/x^{2}$ potential has been proposed as a candidate
for describing a black hole.
Tachyon equations of motion have been obtained,
which look like
the correct ones for a black hole. This has been the basis of
the identification.
However, in two dimensions,
coordinate transformations make it possible to remove any trace
of the metric from the wave equation. In fact, this is achieved
by using precisely the
time-of-flight coordinate. Clearly there is no
information about the black hole in the wave equation unless it
is supplemented by information about the gauge or equivalently
about the dilaton. This makes this line of approach difficult.

Others have tried to make {\it only} a field transform, without any change
of the underlying matrix model \cite{wad,das}.
This, typically, suggests that the
cosmological constant should be identified as the black-hole mass.
This cannot, however, lead to the correct scaling, as we will see.

In \cite{jev} it was suggested that one needs to have {\it both}
a new matrix model, modified by a $1/x^{2}$ piece,
and a field transformation. In such a way
the authors
were able to show that the resulting model has a set of desirable
properties, not shared by above attempts.
In this paper I will examine the modified matrix model further. I
will use standard $c=1$ matrix-model tools to redo some calculations
in the modified model. These results should be useful in the
further study of this model, which might be the long searched for
matrix model black hole.

\section{A Modified Matrix Model}

In $c=1$ string theory, there exists an infinite set of special
(or discrete) states apart from the tachyon, the only field degree
of freedom. These states can be excited by operators,
which are constructed, in the matrix model,
by the eigenvalue coordinate $x$ and its
conjugate momentum $p$. They all have counterparts in field theory, as
discussed in \cite{new,min1,min2}. The field theory special states
fall into two subsets, the Seiberg (S) and
anti-Seiberg (AS), \cite{sei,shenk}.
An example of an AS operator
is the operator that deforms the flat dilaton vacuum into
a black hole, \cite{kuta}.
The matrix model special states correspond to the S states, and
their identification is fairly well understood. So far,
however, the issue of the AS states in the matrix model has been very
confusing. For some possible clues see \cite{wit2}.

The operators all have their particular gravitational scaling. This is
an important clue for the correct identification. Can we find an object
in the matrix model with the same scaling as the black-hole operator
discussed above? Indeed we can: the simplest choice is $1/x^{2}$.
What is remarkable with this potential, a property
that it shares with the pure
$x^{2}$, is that it is exactly soluble. Even in this modified potential
the energy levels are equally spaced (if the $x^{2}$ is continued to
the right-side-up version, see figure 1). This is a property not shared
by a generic potential. As discussed in \cite{min1}, it is the resonances
in the continued potential that give rise to the spectrum of special
states. Therefore, the equal spacing is an important property of the
potential if it is to be a candidate for the black hole.

One must also verify that the correct scaling is obtained. In particular
we would like anomalous logarithms. These indicate that we have
an extra spatial dimension. Again this is a property not shared by
many potentials, see \cite{avhand}. But, as explained in \cite{jev}, with
$1/x^{2}$, it works. To understand how, take a look at figure 2.
The Fermi sea is supposed to reach precisely
to the top of the undeformed $-x^{2}$ part of the potential. The
critical point, then, is approached as the $\eta /x^{2}$
deformation is removed, i.e. $\eta \rightarrow 0$. The double scaling
limit is obtained by keeping the string coupling constant
$1/(\beta \eta ^{1/2})$ fixed. $\beta$ is $1/\hbar$ and goes to infinity
in this limit. Throughout the paper all $\beta$'s will be absorbed
into $\eta$ when appropriate.

\section{Wave Functions}

In this section I will solve the Schr\"{o}dinger equation for the
modified potential.
The equation is
\be
\frac{1}{2} y'' +\left( \frac{1}{2\alpha '} x^{2} -
\frac{\eta}{x^{2}}\right) y
=-Ey    .
\ee
I will, as in \cite{min1,min2}, make the continuation
$\alpha ' \rightarrow -\alpha '$, figure 1, and solve
\be
y'' + \left( -x^{2} +\epsilon  -
\frac{2\eta}{x^{2}} \right) y=0
\ee
instead; $x$ has been rescaled and
$\epsilon = 2i(\alpha ')^{1/2} E$.
With
\be
2\eta = a^{2}-(-1)^{n}a     \label{eta}
\ee
one finds the energy levels
\be
\epsilon _{n} = 2n+1+2a      .
\ee
The normalized wave functions are
\be
\frac{1}{\left[ k! \Gamma (a +\frac{1}{2}+k) \right] ^{1/2}}
H_{2k}^{(a)}(x)|x|^{a}e^{-\frac{x^{2}}{2}}  \label{evenw}
\ee
for $n=2k$ even and
\be
\frac{1}{\left[ k! \Gamma (a +\frac{3}{2}+k) \right] ^{1/2}}
H_{2k+1}^{(a)}(x)|x|^{a}e^{-\frac{x^{2}}{2}}  \label{oddw}
\ee
for $n=2k+1$ odd. The polynomials $H_{n}^{(a)}$ are modifications of the
Hermite polynomials that are recovered at $a=0$.
They can be written in terms of degenerate hypergeometric functions
\be
H_{2k}^{(a)}(x) =(-1)^{k} (a+\frac{1}{2}) _{k}
\;_{1} F_{1} (-k
;a+\frac{1}{2};x^{2})
\ee
for $n=2k$ even and
\be
H_{2k+1}^{(a)}(x) =(-1)^{k} (a+\frac{3}{2})_{k} \; x_1 F_{1} (-k
;a+\frac{3}{2};x^{2})
\ee
for $n=2k+1$ odd.
$(x)_{k} = x(x+1)...(x+k-1)$ is the Pochhammer symbol.
The modified Hermite polynomials
obey a recursion relation
\be
H_{n+1}^{(a)} = xH_{n}^{(a)} - \lambda _{n} H_{n-1}^{(a)},\label{rec}
\ee
where
\be
\lambda _{n} = \left\{
\begin{array}{c}
a+ \frac{n}{2} \;\;\; n\; \mbox{odd} \\
\frac{n}{2} \;\;\;\;\;
n\;\mbox{even}
\end{array}
\right.      .
\ee
Because of (\ref{eta}), the value of $a$ will be
different for odd and even $n$.
An appropriate $\eta =0$ limit gives
\be
a= -\frac{1}{2} +\sqrt{\frac{1}{4}+2\eta}
\ee
for $n$ odd and the negative of this for $n$ even. This means that
the set of wave functions (\ref{evenw},\ref{oddw})
is doubled. One for each sign of
$a$. There are also different recursion relations for the different
signs. The {\it physical} wave functions are just half of these,
the odd ones of the positive-sign set and the even ones of the
negative-sign set (if $\eta >0$). The unphysical wave functions
are still of importance as intermediate steps in the recursion relations.

An important tool for the study of the harmonic oscillator is the
step up and down operators which connects different energy levels.
Its existence is a consequence of the equal spacing of the
energy levels. Since the situation is similar here, we might expect
the existence of such operators here as well. However, the equal
spacing refers to the even and odd levels separately only. Therefore
we may only construct operators that take two steps at a time.
The operators are
\be
b=\frac{1}{2}(ip+x)^{2} -\frac{\eta}{x^{2}},
\ee
its conjugate $b^{\dagger}$ and the hamiltonian
\be
H= \frac{1}{2}(p^{2}+x^{2})+\frac{\eta}{x^{2}}   .
\ee
They obey
$$
\left[ b, b^{\dagger} \right] =4H      ,
$$
\be
\left[ H,b\right]=-2b  \;\;\; \mbox{and}
\;\;\; \left[ H,b^{\dagger} \right]
=2b^{\dagger}                    .
\ee
The algebra is the same as the one of $a^{2}$, $a^{\dagger 2}$, and
$H$ in the case of the harmonic oscillator. There also exist operators
$\frac{ip+x}{\sqrt{2}} \pm \frac{\eta ^{1/2}}{x}$
which connect the even and odd states. They differ from the
$b$'s and $b^{\dagger}$'s in the sense that if you act on
an arbitrary energy eigenstate, you will obtain a new eigenstate
only in half of the cases.

It is the harmonic oscillator algebra in the undeformed model
that lies behind the $W_{\infty}$ structure in the matrix
model. Since a $W_{\infty}$ is expected also in the black hole
case, see e.g. \cite{jackd}, it is a necessary requirement of
the modified model that it
exhibits a similar structure, as indeed it does.

Since the potential is singular we must be careful and check if the
wave functions we use
are normalizable. I said above that the expressions
(\ref{evenw}) and (\ref{oddw}) were of unit norm.
More precisely, they are of unit norm
{\it or} not normalizable at all.
The wave functions that are in danger of being unnormalizable are
the even ones. In fact, when $2\eta > \frac{3}{4}$ they all blow up
worse than $1/\sqrt{x}$ at $x=0$, are therefore not normalizable and
hence must be discarded. This is a very desirable property. In fact,
according to \cite{jev,dvv},
the poles of the tachyon correlation functions
in the black hole have a spacing that is twice as large as the one
in the flat dilaton theory. In units where $\alpha ' =1$ they occur
only at even integer momenta rather than at all integer momenta.
The matrix model explains this in a neat
way in the weak coupling regime
($\eta$ large and positive)
by throwing out the even states. The poles are due to
transitions between states in the continued oscillator \cite{min1},
which will then have twice the spacing.

When $-\frac{1}{4}<2\eta < \frac{3}{4}$, i.e.
at strong coupling,
all wave functions are normalizable and
we have
a choice of which ones to keep.
Let us consider this in more detail. Owing to the
singular potential it is reasonable to try to limit the
theory to just one side of the singular point. Nothing is allowed to pass
through it. To keep just the odd wave functions is a way to do this
consistently. In the case of strong coupling, however,
there is a one-parameter family of different
ways to do this, i.e. different
self-adjoint extensions \cite{adj}. These are determined by
the requirement that the kinetic operator, $\frac{d^{2}}{dx^{2}}$, must
be self-adjoint. From
\be
\int ^{0} \psi ^{\dagger} \frac{d^{2}\psi}{dx^{2}}  =  \int ^{0}
\frac{d^{2}\psi ^{\dagger}}{dx^{2}} \psi
+\left.
\left( \psi^{\dagger}\frac{d\psi}{dx}
-\frac{d\psi ^{\dagger}}{dx}\psi \right) \right|
_{x=0}
\ee
it follows that
\be
\left. \left(
\psi^{\dagger}\frac{d\psi}{dx}
-\frac{d\psi ^{\dagger}}{dx}\psi \right) \right| _{x=0} =0 .\label{prob}
\ee
Not all wave functions obey this requirement, but those built on a basis
consisting of
$$
\psi _{r,\epsilon} \sim rx^{-a} e^{-\frac{x^{2}}{2}} \; _{1}
F_{1}(-\frac{\epsilon}{4} -\frac{a}{2} +\frac{1}{4} ;-a+\frac{1}{2};
x^{2})
$$
\be
+x^{a+1} e^{-\frac{x^{2}}{2}} \; _{1}
F_{1}(-\frac{\epsilon}{4} +\frac{a}{2} +\frac{3}{4} ;a+\frac{3}{2};
x^{2})   \label{sae}
\ee
do; $r$ is a free parameter (independent of $\epsilon$)
labelling the self-adjoint extension.
It is
easy to check that any complex linear combination of wave functions
(\ref{sae}) with different energies $\epsilon$ obeys (\ref{prob}).
Note that I have not fixed the energy levels $\epsilon$ yet.
{}From (\ref{sae}) one recognizes a combination of
(\ref{evenw}) and (\ref{oddw}).
The energy levels are determined by requiring the wave functions
to vanish asymptotically.
This leads to the following condition
\be
r\frac{\Gamma (-a+\frac{1}{2})}
{\Gamma (-\frac{\epsilon}{4} -\frac{a}{2} +\frac{1}{4})} +
\frac{\Gamma (a+\frac{3}{2})}
{\Gamma (-\frac{\epsilon}{4} +\frac{a}{2} +\frac{3}{4})} =0 .
\ee
At $r=0$ we recover the odd wave functions and their
equally-spaced energy levels.
At general $r$, things are much more
complicated and the energy levels are no longer equally spaced.
In \cite{jev} it is proposed that $\eta <0$
correspond to a negative mass, and hence naked, singularity.
The strong coupling ambiguity we have discussed is relevant for both
strongly coupled black holes and strongly coupled naked singularities.

The case of a weakly coupled naked singularity is however very
different.
When $2\eta <-\frac{1}{4}$
the energy eigenvalues turn complex, which signals an
instability where everything rushes
into the infinite $-1/x^{2}$ potential well.

Instead of using the terms weak and strong coupling, one might
talk about large- and small-mass black holes. It then follows
that the ambiguity exists for small mass black holes, or
naked singularities of small negative mass. Furthermore, naked
singularities of large negative mass are not allowed.
This behaviour of the matrix model in the presence of a naked
singularity, if the interpretation of
\cite{jev} is correct, is very interesting and might have important
consequences.

In the following I will keep only the odd wave functions and concentrate
on the weak-coupling, positive-mass, black hole.

\section{The Puncture Two-Point Function}

The first object I will calculate is the two-puncture
correlation function. This is an elementary exercise.
A puncture is a zero-momentum tachyon,
which may be inserted by taking a derivative with respect
to the cosmological constant. The two-point correlation function
is given by \cite{nick}:
\be
\langle PP \rangle
= -\frac{1}{\pi} \Im \sum _{n=0}^{\infty} \frac{1}{E_{n}- \mu} .
\ee

In our case, $\mu =0$ and the sum should be only over odd $n$, i.e.
\be
-\frac{\sqrt{\alpha '}
}{2\pi} \sum _{m=0}^{\infty}
\frac{1}{m+\frac{1}{2}+\frac{z}{2}} ,
\ee
where $z=\frac{1}{2}+a = \sqrt{2\eta +\frac{1}{4}}$.
Now use
\be
\sum _{n=0}^{\infty} \frac{1}{n+x+y} =\psi (x+y)
= B_{0}(x)\log y + \sum _{n=1}^{\infty} B_{n}(x)(-1)^{n}
\frac{-1}{ny^{n}}         ,
\ee
where $B_{n}(x)$ are the Bernoulli polynomials.
The sum is not convergent, but can be
defined by taking a $y$ derivative
and throwing away terms analytic in $y$.
The result is
\be
\langle PP \rangle
=-\frac{\sqrt{\alpha '}}{2\pi}\left(
B_{0} (1/2) \log z -
\sum _{n=1}^{\infty}
B_{2n}(1/2) \frac{2^{2n-1}}{nz^{2n}}  \right) .
\ee
I have used that $B_{k}(1/2)=-(1-2^{1-k})B_{k} =0$
for odd $k$ to get rid of the odd powers. The Bernoulli numbers
$B_{k}$ are zero for all odd $k$ except $k=1$.
The expansion in terms of $z$ is,
apart from an alternating
sign, identical to the one of $1/\mu ^{2}$ in the
unmodified potential. However, the above expression must be reexpanded
in $1/\eta$, which is the string coupling squared.
Up to genus two the result is
\be
-\frac{\sqrt{\alpha '}}{2\pi} \left( \log \eta ^{1/2} +
\frac{7}{48 \eta} - \frac{167}{3840\eta ^{2}} +... \right)
\ee
compared to
\be
-\frac{\sqrt{\alpha '}}{\pi} \left( \log \mu -
\frac{1}{24 (\sqrt{\alpha '} \mu )^{2}} -
\frac{7}{960(\sqrt{\alpha '} \mu )^{4}} -... \right)
\ee
for the undeformed case.

\section{Tachyon Correlation Functions}

In this section I will calculate some correlation functions at
non-zero momentum, culminating in expressions for tachyon correlation
functions up to genus two.

We first need the recursion relations (\ref{rec}) expressed in terms of
normalized wave functions. They are
\be
\left( \frac{n+1}{2}\right) ^{1/2} | n+1 \rangle =x| n \rangle -
\left( \frac{n}{2} +a\right) ^{1/2} |n-1 \rangle
\ee
for $n$ odd and
\be
\left( \frac{n+1}{2}+a\right) ^{1/2} |n+1 \rangle =x|n \rangle-
\left( \frac{n}{2} \right) ^{1/2} |n-1 \rangle
\ee
for $n$ even. Note that it is only those
$|m \rangle$ where $m$ is odd that are {\it physical}.

With the help of these recursion relations, it is simple to derive
many matrix elements. For instance
\be
\langle n | x^{2k} |n+2k\rangle =
\left[ (\frac{n}{2}+a+1)_{k} (\frac{n+1}{2})_{k} \right] ^{1/2} .
\label{mat}
\ee
Correlation functions of $x^{2k}$ will have poles for a set of
different momenta, see \cite{min1} for the undeformed case.
The highest-momentum pole is due
to the tachyon piece of $x^{2k}$. In terms
of the operators $b$ and $b^{\dagger}$ we have
\be
x^{2k} = 2^{k-1} (b^{k} + b^{\dagger k}) +...
\ee
To calculate two-point tachyon correlation functions we therefore
only need the matrix element (\ref{mat}). The starting point is
\be
\langle PPO_{1}O_{2} \rangle
=\frac{1}{\pi} \Im \sum _{n,m=0}^{\infty}
\frac{\langle n|O_{1}|m\rangle
\langle m|O_{2}|n\rangle
}{(E_{n}-\mu )^{2}} \frac{2(E_{m}-E_{n})}
{p^{2}+(E_{m}-E_{n})^{2}}            ,
\ee
which was derived in \cite{min1} using perturbation theory.
For comparison I will first do the calculation for the undeformed
model. Calculations using a similar method have been done in
\cite{min2}, and in particular in \cite{avhand} up to genus seven.
In this case we find
\be
<PT_{p}T_{p}> = -\frac{2^{-2k+1}}{\pi}
\Re \sum_{n=0}^{\infty}
\frac{(n+1)_{2k}-(n)_{-2k}}{2n+1+2i\sqrt{\alpha '}\mu }
\frac{4k}{p^{2}-\frac{4k^{2}}{\alpha '}} .  \label{stand}
\ee
The sum is defined by taking enough derivatives with respect to $\mu$
in order for it to be convergent. Analytic terms are then skipped.
With this prescription it follows that
\be
\sum _{n=0}^{\infty} \frac{n^{k}}{2n+1+2i\sqrt{\alpha '}\mu}
= \left( -\frac{1+2i\sqrt{\alpha '}\mu}{2}\right) ^{k}
\sum _{n=0}^{\infty} \frac{1}{2n+1+2i\sqrt{\alpha '}\mu}. \label{sum}
\ee
The difference in the numerator of (\ref{stand}) can be expanded in
large $n$ and $\mu$ (this will eventually give the genus expansion)
using Stirling numbers of the first kind.
They are defined by
\be
(x)_{-n} =\sum_{m=0}^{n} S_{n}^{(m)} x^{m}  ,
\ee
where the first few Stirling numbers are

\newpage

$$
S_{n}^{(n)}=1,\;\;S_{n}^{(n-1)} = -\left( \begin{array}{c} n\\2
\end{array} \right),\;\; S_{n}^{(n-2)} = \left( \begin{array}{c}
n\\3 \end{array} \right) \frac{3n-1}{4},
$$
$$
S_{n}^{(n-3)} =
-\left( \begin{array}{c} n\\4 \end{array} \right) \frac{n(n-1)}{2},\;\;
S_{n}^{(n-4)} =
\left( \begin{array}{c} n\\5 \end{array} \right)
\frac{15 n^{3} -30 n^{2} +5n+2}{48}
$$
\be
\mbox{and} \;\;
S_{n}^{(n-5)} =
-\left( \begin{array}{c} n\\6 \end{array} \right)
\frac{n(n-1)(3n^{2} -7n -2)}{16}         .   \label{sterling}
\ee
The first two are given in \cite{abr}, while the others can be
calculated from relations given in \cite{abr}.
The numerator is then
\be
2^{-2k} \sum _{m=0} ^{2k} \left[ S_{2k+1}^{(m+1)} (-1)^{m}-
S_{2k}^{(m)} \right] n^{m}            .
\ee
With the help of (\ref{sum}) and (\ref{sterling})  the
correlation function up to genus two can
be shown to be
$$
\frac{\sqrt{\alpha '} 2^{-2k+5}}{\pi}
k^{3}\left[ (\sqrt{\alpha '}\mu )^{2k-1}
-\frac{1}{12} (2k-1)(k-1) (4k^{2}-2k-1)
(\sqrt{\alpha '}\mu )^{2k-3}  \right.
$$
$$
\left.
+\frac{(2k-1)(k-1)(2k-3)(k-2)}{1440} (48k^{4}-80k^{3}-20k^{2}+24k+7)
(\sqrt{\alpha '}\mu )^{2k-5}
\right]
$$
\be
\times
\frac{\log \mu}{p^{2}-\frac{4k^{2}}{\alpha '}}  .
\ee
The calculation is quite tedious for higher genus. {\it Mathematica}
is recommended from genus two.
As discussed in \cite{min1,min2} this should
be interpreted as
$$
\frac{2^{-2k+2}}{\pi}
p^{3}\left[ \mu ^{p-1}
-\frac{1}{24} (p-1)(p-2) (p^{2}-p-1)
\mu ^{p-3}  \right.
$$
$$
\left.
+\frac{(p-1)(p-2)(p-3)(p-4)}{5760}(3p^{4}-10p^{3}-5p^{2}+12p+7)
\mu ^{p-5}
\right]
$$
\be
\times
\frac{\log \mu}{p^{2}-4k^{2}}        \label{van}
\ee
in agreement with \cite{moore} for $\alpha ' =1$.
The pole factor is the leading part of the
gamma function external legs.

Let me repeat the calculation, now
for the deformed model at $\mu =0$. We then have
$$
\langle PPT_{p}T_{p}\rangle
=
-\frac{4\sqrt{\alpha '}}{\pi}
\sum_{m=0}^{\infty}
\frac{(m+1+z)_{k}(m+1)_{k}-(m+z)_{-k}(m)_{-k}}{(2m+1+z)^{2}}
$$
\be
\times \frac{4k}{p^{2}-\frac{4k^{2}}{\alpha '}}   .
\ee
The numerator is
\be
\sum _{i,j=0} ^{k} \left[ S_{k}^{(i)} S_{k+1}^{(j+1)} (-1)^{i+j}
-S_{k+1}^{(i+1)}S_{k}^{(j)} \right] (m+1+z)^{i}m^{j}. \label{num2}
\ee
The expression is divisible by $2m+1+z$ once. To see this, note that
(\ref{num2}) is a sum of terms each proportional to $(m+1+z)^{j-i}-
(-1)^{i+j}m^{j-i}$ for some $j>i$. This is zero for $1+z=-2m$.
This is why it is
convenient to consider an extra puncture in the correlation
function as compared to (\ref{stand}).
The correlation function is then
$$
\frac{\sqrt{\alpha '}
(-1)^{k} 2^{-2k+6}}{\pi}
k^{3}\left[
z^{2k-2} - \frac{1}{3}(k-1)
(4k^{2}-2k-3)z^{2k-4} \right.
$$
\be
\left.
+\frac{(k-1)(k-2)}{90}(80k^{4}-144k^{3}-152k^{2}+132k+105) z^{2k-6}
\right]
\frac{\log z}{p^{2}-\frac{4k^{2}}{\alpha '}}    \label{zutr}
\ee
up to genus two. In terms of $\eta$:
$$
\frac{(-1)^{k} 2^{-k+2}}{\pi} p^{3}
\left[
\eta ^{p/2-1} - \frac{1}{12}(p-2)
\left( p^{2}-p-\frac{15}{4}\right)
\eta ^{p/2-2}
\right.
$$
\be
\left.
+\frac{(p-2)(p-4)}{23040}(80p^{4}-288p^{3}-728p^{2}+
1176p + 2085) \eta ^{p/2-3}
\right]
\frac{\log \eta
}{p^{2}-4k^{2}}    .     \label{svart}
\ee
at $\alpha ' =1$.
This is the final answer for the tachyon two-point function in
the Euclidean black hole to genus
two and the most important result of the paper.
Expressions with fewer or more punctures are easily obtained by
integrating or taking derivatives with respect to the
$z$ in the denominator of (37)
and then reexpanding in $1/\eta$.
The method I have used produces the collective field theory part of
the amplitude unambiguously. However, the only information it
gives about the factorized external legs is where the poles are.
This is so because it is valid only for the discrete tachyons which
are sitting right on the poles.
In the undeformed case of the formula, (\ref{van}),
$k$ may be a half-integer,
in the black-hole case of
(\ref{svart}), only integers are allowed.
It would therefore be desirable to redo the calculation using
the method of \cite{moore}. I hope to return to this issue in the
future.

\section{Conclusions}

As indicated in this paper, correlation functions in the $1/x^{2}$
deformed matrix model are easily calculated using methods similar
to the ones used
for the standard $c=1$ matrix model. The results are different,
but with a similar structure.
It is also interesting to note that there is a restriction on possible
naked singularities: they can not be of arbitrary negative mass.

An important task that remains is an
{\it explicit} derivation of the black hole from the modified
matrix model. In the standard $c=1$ such a derivation exists,
i.e. the loop operator construction, where the Liouville mode
is clearly identified in the matrix model. Such a construction
is necessary also for the black hole if the identification
proposed in \cite{jev} is to be established.

It is important to realize that the $1/x^{2}$ deformed model
is a candidate for an {\it eternal} black hole. As such it may not
have much to say about the information problem, which is best discussed
in the context of forming and evaporating black holes. On the
other hand, with a matrix-model black hole at hand, even if it is
eternal, we might learn to recognize matrix-model black holes in
general. This should help us towards a full understanding of
what matrix models, i.e. string theory, have to tell us about
black holes.

\section*{Acknowledgements}

I would like to thank David Gross for discussions.

\end{document}